\documentclass[runningheads]{llncs}

\usepackage{eccv}
\usepackage{eccvabbrv}
\usepackage{graphicx}
\usepackage{booktabs}
\usepackage[export]{adjustbox}
\usepackage{tikz}
\usepackage{algorithm}
\usepackage{algpseudocode}
\usepackage[accsupp]{axessibility}
\usepackage{hyperref}
\usepackage{orcidlink}

\begin{document}

\title{Lossy Image Compression with Foundation Diffusion Models} 

\author{
Lucas Relic\inst{1,2}
\and Roberto Azevedo\inst{2}
\and Markus Gross\inst{1,2}
\and Christopher Schroers\inst{2}
}
\authorrunning{L.~Relic et al.}
\institute{ETH Z\"urich, Switzerland \\
\email{\{lucas.relic, grossm\}@inf.ethz.ch}
\and
Disney Research | Studios, Z\"urich, Switzerland \\
\email{\{roberto.azevedo, christopher.schroers\}@disneyresearch.com}
}

\maketitle

\begin{abstract}
Incorporating diffusion models in the image compression domain has the potential to produce realistic and detailed reconstructions, especially at extremely low bitrates.
Previous methods focus on using diffusion models as expressive decoders robust to quantization errors in the conditioning signals.
However, achieving competitive results in this manner requires costly training of the diffusion model and long inference times due to the iterative generative process.
In this work we formulate the removal of quantization error as a denoising task, using diffusion to recover lost information in the transmitted image latent.
Our approach allows us to perform less than 10\% of the full diffusion generative process and requires no architectural changes to the diffusion model, enabling the use of foundation models as a strong prior without additional fine tuning of the backbone.
Our proposed codec outperforms previous methods in quantitative realism metrics, and we verify that our reconstructions are qualitatively preferred by end users, even when other methods use twice the bitrate.
\end{abstract}

\section{Introduction}
\label{sec:intro}

\begin{figure}[t]
    \centering
    \includegraphics[width=0.99\linewidth]{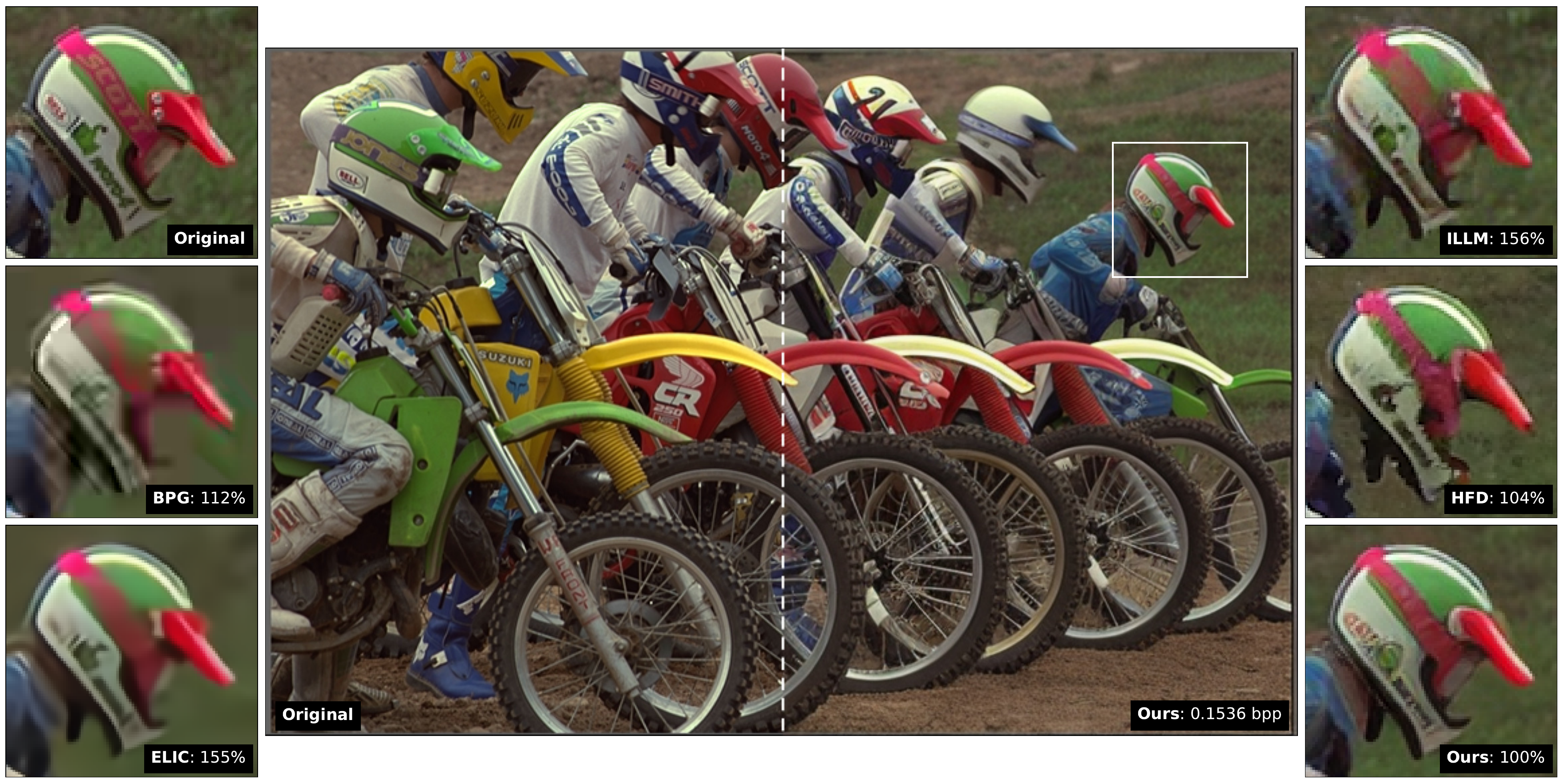}
    \caption{Visual examples of our proposed method and various classes of image compression codecs. Traditional~(BPG~\cite{BPG}) and autoencoder-based~(ELIC~\cite{he2022elic}) codecs suffer from blocking or blurring, and reconstructions from GAN-based~(ILLM~\cite{muckley2023Improving}) and previous diffusion-based~(HFD~\cite{hoogeboom2023HighFidelity}) methods contain high-frequency artifacts. Our proposal is as realistic as the original image while recovering a high level of detail. Bitrates are shown relative to our method. Best viewed digitally.}
    \label{fig:teaser}
\end{figure}

In today's digital era, multimedia content dominates global internet traffic, making the development of efficient compression algorithms increasingly important.
Traditional codecs, which use handcrafted transformations~\cite{wallace1992jpeg, BPG}, are now outperformed by data-driven neural image compression~(NIC)~\cite{balle2017Endtoend,balle2018Variational,minnen2018Joint} methods that optimize for both rate and distortion.
Nevertheless, most current methods still produce blurry and unrealistic images in extremely low bitrate settings~\cite{hoogeboom2023HighFidelity,careil2023image,agustsson2019Generative}.
This is the result of such methods being optimized for rate-distortion, where distortion is measured with pixel-wise metrics like mean-squared error~(MSE)~\cite{blau2019Rethinking}.
The rate-distortion-realism\footnote{We use ``realism'' and ``perception'' interchangeably, representing the similarity of the reconstructed image to other natural images.} triple tradeoff~\cite{blau2019Rethinking, agustsson2023MultiRealism, mentzer2020HighFidelity} formalizes this phenomenon and states that optimizing for low distortion~(\ie, pixel-wise error) \emph{necessarily} results in unrealistic images~(\ie, images that do not fall on the manifold of natural images).
In low-bitrate scenarios, however, it can be preferable to decode realistic~(thus, more perceptually pleasant) images, even if that means lower performance in pixel-wise metrics~\cite{bachard2024coclico}.

Generative compression methods~\cite{mentzer2020HighFidelity, muckley2023Improving, agustsson2019Generative} try to  reconstruct realistic images by introducing GAN architectures and adversarial or perceptual losses.
In image generation, however, diffusion models~\cite{dhariwal2021Diffusion} have now emerged as a powerful alternative, outperforming GANs~\cite{dhariwal2021Diffusion} and achieving state-of-the-art realism scores~\cite{rombach2022HighResolution}.
Diffusion models are thus a natural fit for generative image compression architectures targeting low-bitrate scenarios.
Yet, their applicability and adoption is hindered by large model size and prohibitively expensive training times, requiring multiple GPU \emph{years} and hundreds of thousands of dollars~\cite{chen2023PixArt}.
The introduction of open-source foundation models~\cite{bommasani2021opportunities} has the potential to democratize these powerful models and provide strong priors that can be explored for feature extraction or transfer learning on a variety of domains~\cite{graikos2022Diffusion}, for example image generation~\cite{zhang2023Adding}, depth estimation~\cite{ke2023repurposing}, and even music generation~\cite{forsgren2022Riffusion}.

The use of foundation diffusion models as prior for image compression, however, is still an underexplored research area.
Some works address this task~\cite{lei2023Text, bordin2023semantic} but operate at extremely low bitrates~(less than 0.03 bpp) where reconstructed image content significantly differs from the original, limiting applicability.
Only Careil~\etal~\cite{careil2023image} apply foundation diffusion models in a practical compression setting.
However, they modify the base model architecture and thus require fine-tuning on a large dataset containing millions of images.
Other works, which train the diffusion component from scratch,  operate at relatively high bitrates~\cite{yang2023Lossy} where low pixel-wise distortion can be achieved, or perform image enhancement~\cite{hoogeboom2023HighFidelity, goose2023Neural} rather than native end-to-end compression.
Notably, all current work on diffusion-based image compression tasks sample the output image from pure noise, requiring the full diffusion sampling process, which can take up to one minute per image~\cite{yang2023Lossy} due to its iterative nature.

To advance the state of the art, we propose a novel image compression codec that uses foundation latent diffusion models as a means to synthesize lost details, particularly at low bitrate.
Leveraging the similarities between quantization error and noise~\cite{balle2017Endtoend}, we transmit a quantized image latent and perform a subset of denoising steps at the receiver corresponding to the noise level~(\ie, quantization error) of the latent~(similar to diffusion image editing techniques~\cite{meng2022SDEdit}).
The key components of our proposal are:
i)~the autoencoder from a foundation latent diffusion model to transform an input image to a lower-dimensional latent space;
ii)~a learned \emph{adaptive quantization} and \emph{entropy encoder}, enabling inference-time control over bitrate within a single model;
iii)~a learned method to \emph{predict the ideal denoising timestep}, which allows for balancing between transmission cost and reconstruction quality;
and iv)~a diffusion decoding process to synthesize information lost during quantization.
Unlike previous work, our formulation requires only a fraction of iterative diffusion steps and can be trained on a dataset of fewer than 100k images.
We also directly optimize a distortion objective between input and reconstructed images, enforcing coherency to the input image while maintaining highly realistic reconstructions~(Fig.~\ref{fig:teaser}) due to the diffusion backbone.

In sum, our contributions are:
\begin{itemize}
    \item We propose a novel latent diffusion-based lossy image compression pipeline that is able to produce highly realistic and detailed image reconstructions at low bitrates.
    \item To achieve this, we introduce a novel parameter estimation module that simultaneously learns \emph{adaptive quantization parameters} as well as the \emph{ideal number of denoising diffusion steps}, allowing a faithful and realistic reconstruction for a range of target bitrates with a single model.
    \item We extensively evaluate state-of-the-art generative compression methods on several datasets via both objective metrics and a user study.
    To the best of our knowledge, this is the first user study that compares generative diffusion models for image compression.
    Our experiments verify that our method achieves state-of-the-art visual quality as measured in FID and end users subjectively prefer our reconstructions.
\end{itemize}

\section{Related Work}
\label{sec:related}

Although diffusion models have seen significant successes in the machine learning community, their use in the image compression domain is still limited.

Yang and Mandt~\cite{yang2023Lossy} proposed the first transform-coding-based lossy compression codec using diffusion models.
They condition a diffusion model on contextual latent variables produced with a VAE-style encoder.
Despite showing competitive testing results, their method operates in a relatively high bitrate range~(0.2bpp and above).
It thus leaves room for improvement, particularly at lower bitrates, where the powerful generation capabilities of diffusion models can be used to reconstruct images from lower entropy signals.

Diffusion models have also been proposed to augment existing compression architectures by adding details to images compressed with autoencoder-based neural codecs~\cite{hoogeboom2023HighFidelity, goose2023Neural}.
These methods are sub-optimal since they address an image enhancement task decoupled from compression; in other words, they post-process a compressed image rather than train an image compression method end-to-end.
Images reconstructed in this manner often contain high-frequency artifacts or entirely lose image content as the diffusion model cannot rectify artifacts introduced in the initial compression stage.

Several works develop codecs for extremely low bitrate compression~(less than 0.03 bpp) via latent diffusion.
These methods condition pretrained text-to-image diffusion models with text and spatial conditioning such as CLIP embeddings~\cite{lei2023Text, bachard2024coclico} and edge~\cite{lei2023Text} or color~\cite{bachard2024coclico} maps, respectively.
Most notably, Careil~\etal~\cite{careil2023image} augment a latent diffusion model with an additional encoder and image captioner to produce vector-quantized ``hyper-latents'' and text signals which are used to generate an image at the receiver.
However, their architectural changes require substantial fine-tuning of the underlying diffusion model, hindering the advantages of using a foundation model.
Additionally, in general, the extremely low bitrate of these methods results in reconstructions that, while realistic, vary significantly in content from the original images.

Notably, all existing works focus on regenerating the image at the receiver side from low-entropy conditioning signals, requiring tens or hundreds of costly diffusion sampling steps per image.
Our novel formulation of processing a quantized latent representation allows us to predict and perform the ideal number of denoising diffusion steps, typically between 2 and 7\% of the full process, depending on the bitrate.
Combined with a learned per-content adaptive quantization, we propose the first diffusion-based image compression codec with inference-time bitrate control and demonstrate that our method produces more realistic reconstructions and more faithfully represents the input image compared to previous works on generative image compression~(see Sec.~\ref{sec:experiments}).

\section{Background}
\label{sec:backgroud}

\subsubsection{Neural Image Compression.}
\label{sec:nic}
Lossy neural image codecs~(NIC) are commonly modeled as autoencoders, in which an encoder $\mathcal{E}$ transforms an image $\mathbf{x}$ to a quantized latent $\mathbf{\hat{y}}=\lfloor \mathcal{E}(\mathbf{x}) \rceil $, while a decoder $\mathcal{D}$ reconstructs an approximation of the original image $\mathbf{\hat{x}}=\mathcal{D}(\mathbf{\hat{y}})$.
Based on Shannon's rate-distortion theory~\cite{shannon1948Mathematical}, during training, $\mathcal{E}$ and $\mathcal{D}$ are optimized to minimize the rate-distortion trade-off:

\begin{equation}
\mathcal{L}_{total} = \mathcal{L}_{bits}(\mathbf{\hat{y}}) + \lambda\mathcal{L}_{rec}(\mathbf{x}, \mathbf{\hat{x}}) 
\label{eq:rd-loss}
\end{equation}

where $\mathcal{L}_{rec}$ is a measure of distortion~(commonly MSE), $\mathcal{L}_{bits}(\mathbf{\hat{y}})$ is an estimate of the bitrate needed to store $\mathbf{\hat{y}}$, and $\lambda$ controls the trade-off between rate and distortion.
According to Shannon's theory, $\mathcal{L}_{bits}(\mathbf{\hat{y}}) = -\log_2 P(\mathbf{\hat{y}})$, where $P$ is a probability model of $\mathbf{\hat{y}}$.

As $\mathcal{L}_{rec}$ is often formulated as a pixel-wise difference between images~\cite{blau2019Rethinking}, especially at low bitrate, it can lead to unnatural artifacts in the reconstructed images~(\eg, blurring).
In such scenarios, it is interesting to design $\mathcal{D}$ such that the distribution of reconstructed images closely follows the distribution of natural images~(\ie, that it produces realistic images), even though this results in lower pixel-wise distortion~\cite{blau2019Rethinking}.
Generative image compression methods, such as our proposal, focus on optimizing this rate-distortion-realism trade-off~\cite{agustsson2023MultiRealism}.

\subsubsection{Diffusion.}

\begin{figure}[t]
    \centering
    \begin{subfigure}{0.45\linewidth}
        \includegraphics[width=0.99\linewidth]{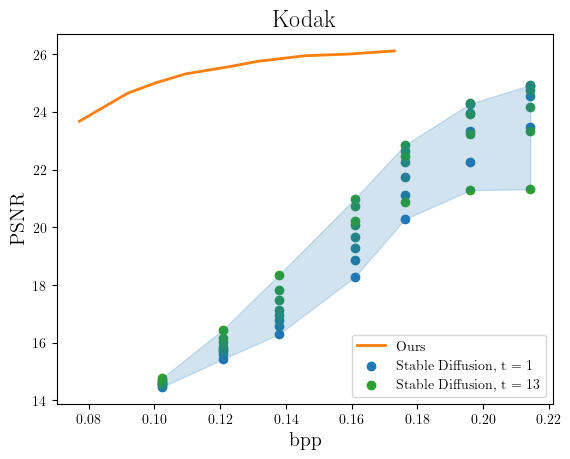}
        \caption{}
        \label{subfig:naive_ldm_rd}
    \end{subfigure}
    \begin{subfigure}{0.5\linewidth}
        \includegraphics[width=0.95\linewidth]{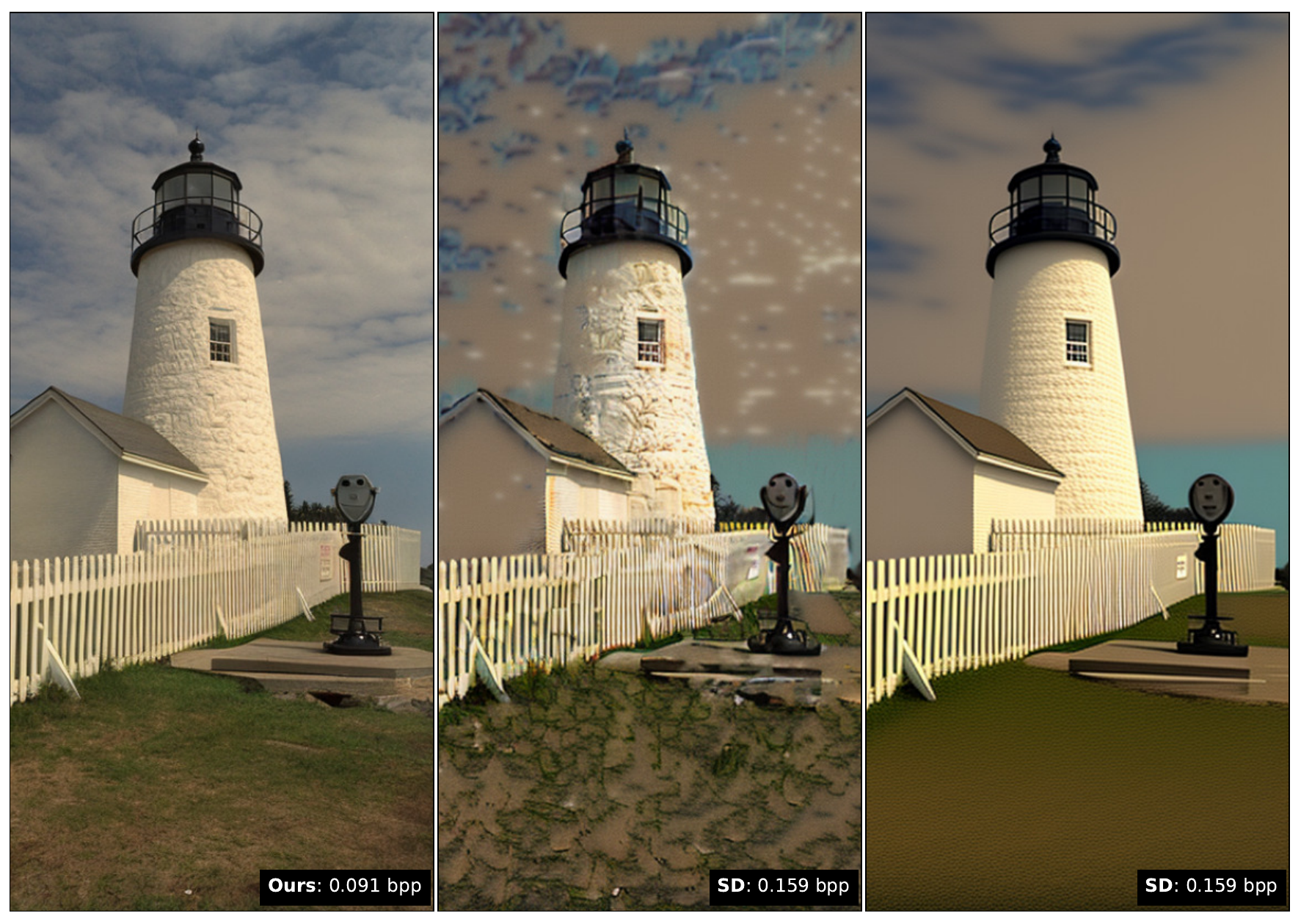}
        \caption{}
        \label{subfig:naive_ldm_vis}
    \end{subfigure}
    \caption{Rate-distortion~(\cref{subfig:naive_ldm_rd}) and visual~(\cref{subfig:naive_ldm_vis}) comparisons of our method to naively quantizing and entropy coding the latents of a latent diffusion model (Stable Diffusion~\cite{rombach2022HighResolution}). The LDM baseline requires nearly triple the bits to achieve comparable performance to our method and severly degrades the image at lower bitrates. Performing additional diffusion steps still does not produce a realistic image (\cref{subfig:naive_ldm_vis}, right). The color gradient of the dots in~\cref{subfig:naive_ldm_rd} represents the number of denoising steps.}
    \label{fig:naive_compression_ldm}
\end{figure}

Diffusion models~(DMs)~\cite{sohl-dickstein2015Deep, ho2020Denoising} are a class of generative models that define an iterative process $q(x_t | x_{t-1})$ that gradually destroys an input signal as $t$ increases, and try to model the reverse process $q(x_{t-1} | x_t)$.
Empirically, the forward process is performed by adding Gaussian noise to the signal; thus, the reverse process becomes a denoising task.
The diffusion model $\mathcal{M}_\theta$ approximates the reverse process by estimating the noise level $\epsilon_\theta$ of the image and using it to predict the previous step of the process:

\begin{gather}
    \mathbf{x}_{t-1} = \sqrt{\alpha_{t-1}}\mathbf{\tilde{x}}_0 + \sqrt{1-\alpha_{t-1}}\epsilon_\theta
    \text{,}\quad\text{with}\quad
    \mathbf{\tilde{x}}_0 = \frac{\mathbf{x}_t - \sqrt{1 - \alpha_t}\epsilon_\theta}{\sqrt{\alpha_t}} \label{eq:denoise}
\end{gather}
Here, $t$ is the current timestep of the diffusion process, $\mathbf{x}_{t}$ and $\alpha_t$ represent the sample and variance of the noise at timestep $t$, respectively, and $\mathbf{\tilde{x}}_0$ is the predicted fully denoised sample from any given $t$.
Eqs.~\ref{eq:denoise} can be simplified to
\begin{equation}
    \mathbf{x}_{t-1} = \mathcal{M}_\theta (\mathbf{x}_t, t)
\label{eq:diffusion}
\end{equation}
where $\mathcal{M}_\theta(\cdot)$ is one forward pass of the diffusion model. It is therefore possible to sample from a DM by initializing $\mathbf{x}_T=\mathcal{N}(0,1)$ and performing $T$ forward passes to produce a fully denoised image.

Latent diffusion models~(LDMs)~\cite{rombach2022HighResolution} improve memory and computational efficiency of DMs by moving the diffusion process to a spatially lower dimensional latent space, encoded by a pre-trained variational autoencoder~(VAE)~\cite{kingma2014auto}.
Such a latent space provides similar performance of the corresponding pixel-space DMs while requiring less parameters~(and memory)~\cite{blattmann2023align}.
These types of DMs are trained in a VAE latent space where $\textbf{y}=\mathcal{E}_{vae}(\textbf{x})$ and a sampled latent $\mathbf{y}_0$ can be decoded back to an image $\mathbf{\hat{x}}=\mathcal{D}_{vae}(\mathbf{y}_0)$.

Since LDMs are based on VAEs, they can also be considered a type of compression method.
However, their applicability in lossy image compression is hindered by inherent challenges.
LDMs lack explicit training to produce discrete representations, resulting in highly distorted reconstructions when used for lossy compression~\cite{hoogeboom2023HighFidelity}, and cannot navigate the rate-distortion tradeoff.
To highlight such issues,~\cref{fig:naive_compression_ldm} shows the performance of the same LDM used by our method~(without modifications) as a compression codec compared to our approach optimized for lossy image compression.
In this experiment, we manually sweep over a range of quantization and diffusion timestep parameters, encoding the images under the different configurations.
Specifically, we encode to the latent space, quantize according to the chosen parameters, compress with zlib~\cite{zlib}, run the chosen number of denoising diffusion steps, and decode back to image space.
As shown, the unmodified LDM requires nearly 3x the bits to achieve comparable performance to our method and cannot produce realistic images at low bitrates, regardless of the number of diffusion steps performed~(\cref{subfig:naive_ldm_vis}).
Thus, deploying LDMs for compression requires thoughtful consideration to maximize their effectiveness.

\section{Method}
\label{sec:method}
\begin{figure}[t]
    \centering
    \includegraphics[width=0.99\linewidth]{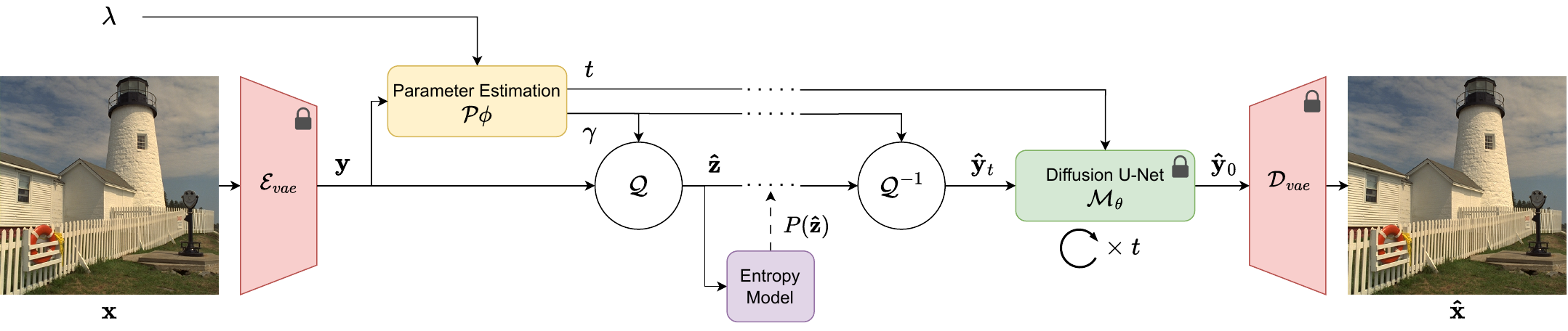}
    \caption{Overview of our approach. The input image $\mathbf{x}$ is encoded into latent space and transformed according to predicted parameters $\gamma$ before quantization and entropy coding. The quantized representation $\mathbf{\hat{z}}$ is transmitted with $\gamma$ and predicted diffusion timestep $t$ as side information. At the reciever the latent is inverse transformed, diffused over $t$ steps, and decoded back to image space.}
    \label{fig:pipeline}
\end{figure}

Fig.~\ref{fig:pipeline} shows the high-level architecture of our method.
It is composed of a variational autoencoder~(containing an encoder, $\mathcal{E}_{vae}$, and a decoder, $\mathcal{D}_{vae}$) a \emph{quantization} and \emph{diffusion timestep} parameter estimation network~($\mathcal{P}_{\phi}$), an entropy model, and a latent diffusion model~($\mathcal{M}_{\theta}$).

Our encoding process is performed as follows:
First, the image $\mathbf{x}$ is encoded into its latent representation $\mathbf{y} = \mathcal{E}_{vae}(\mathbf{x})$.
Then, $\mathbf{y}$ is quantized by an adaptive quantization method parameterized by $\gamma$~(\ie, $\mathbf{\hat{z}} = \mathcal{Q}(\mathbf{y}, \gamma)$).
Finally, $\mathbf{\hat{z}}$ is entropy encoded and stored or transmitted.

During decoding, the inverse quantization transformation computes $\mathbf{\hat{y}}_t = \mathcal{Q}^{-1}(\mathbf{\hat{z}}, \gamma)$, which is then used as input to the generative LDM process over $t$ denoising steps to recover an approximation $\mathbf{\hat{y}}_{0}$ of the original latent representation $\mathbf{y}$.
Finally, $\mathbf{\hat{y}}_{0}$ is decoded by the VAE decoder into a reconstructed image $\mathbf{\hat{x}} = \mathcal{D}_{vae}(\mathbf{\hat{y}_{0}})$.
Algorithm~\ref{alg:inference} shows the complete encoding/decoding process.

A key feature of our method is that both the quantization parameters $\gamma$ and the number of denoising steps $t$ can be adapted in a per-content and per-target-bitrate manner~(controlled by the rate-distortion trade-off parameter $\lambda$).
To achieve this, we train a neural network $\mathcal{P}_{\phi}(\mathbf{y}, \lambda)$ that predicts both $t$ and $\gamma$.

Intuitively, our method learns to discard information~(through the quantization transformation) that can be synthesized during the diffusion process.
Because errors introduced during quantization are similar to adding noise~\cite{balle2017Endtoend, balle2018Variational, minnen2018Joint} and diffusion models are functionally denoising models, they can be used to remove the quantization noise introduced during coding.

\subsection{Latent Diffusion Model Backbone}
Aiming at avoiding extensive training time, we use Stable Diffusion v2.1~\cite{rombach2022HighResolution} for certain modules of our architecture, particularly, $\mathcal{E}_{vae}$, $\mathcal{D}_{vae}$, and $\mathcal{M}_{\theta}$.
Note that our method works independently of the base model.
We select Stable Diffusion as it is one of the only foundation latent diffusion models with publicly available code and model weights.

\subsection{Parameter Estimation}
\label{subsec:param-est}
As aforementioned, the quantization parameters, $\gamma$, and the optimal number of denoising steps the diffusion network should perform, $t$, are predicted by a neural network $\mathcal{P}_\phi(\mathbf{y}, \lambda)$, which takes as input the latent $\textbf{y}$ and the rate-distortion trade-off $\lambda$.

\subsubsection{Adaptive Quantization.}
Our adaptive quantization function $\mathcal{Q}$ is defined as an affine transformation $\mathcal{T}$ for each channel of the latent $\mathbf{y}$, parameterized by $\gamma$, before applying standard integer quantization, \ie,

\begin{equation}
\mathbf{\hat{z}} = Q(\mathbf{y}, \gamma) =  \lfloor \mathcal{T}(\mathbf{y}, \gamma) \rceil
\end{equation}

$\gamma$ is transmitted as side information in order for the decoder to perform the inverse transform at the client side, \ie,
\begin{equation}
\mathbf{\hat{y}}_t = Q^{-1}(\mathbf{\hat{z}}, \gamma) = \mathcal{T}^{-1}(\mathbf{\hat{z}}, \gamma)
\end{equation}

\subsubsection{Timestep Prediction.}
\begin{figure*}[t!]
    \centering
    \includegraphics[width=0.99\linewidth]{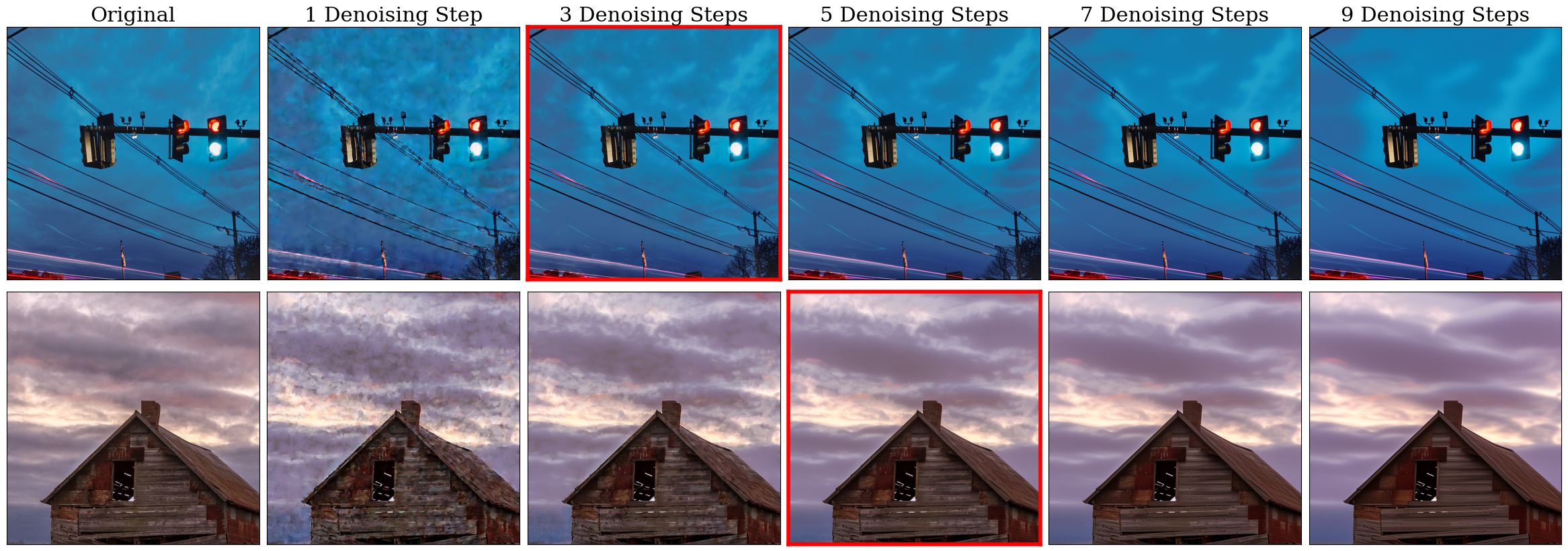}
    \caption{Intermediate states of the sequential denoising process in our decoder. Our method predicts the optimal number of denoising steps, highlighted in red, to produce the most perceptually pleasing output. Best viewed digitally.}
    \label{fig:num_steps}
\end{figure*}

Contrary to an image generation task, which begins diffusion from ``pure noise'', in our compression task, we start the diffusion process from a quantized latent, which already contains structural and semantic information of the content.
In such a scenario, performing the entire range of denoising steps during decoding is both wasteful and results in over-smoothed images.
Therefore, we learn to predict the subset of denoising diffusion steps that produces optimal decoded images.
\cref{fig:num_steps} illustrates how the decoded image quality changes based on the number of diffusion steps performed by the decoder, where too few or too many steps result in noisy or over-smoothed images, respectively.
Because the number of optimal denoising steps depends on the amount of noise in the latent~(and therefore the severity of quantization), and vice versa, we predict $t$ and $\gamma$ jointly in the parameter estimation module.

\subsubsection{Architecture.} 
The parameter estimation neural network $\mathcal{P}_\phi(\textbf{y}, \lambda)$ employs a fully convolutional architecture.
We stack alternating downsampling~(\ie,~stride 2) and standard convolutional layers, increasing filter count as depth increases.
In the last layer, we reduce the output filter count to correspond with the total number of parameters estimated\footnote{We use 2 quantization parameters per latent channel plus one additional parameter for timestep prediction. As the image latent has 4 channels we predict a total of 9 parameters per image.} and apply mean average pooling on each output channel to produce a single scalar for each parameter. 
We use SiLU~\cite{hendrycks2023gaussian} activation between each convolutional layer.
No activation is applied after the final convolutional layer.
However, we apply sigmoid activation on the scalar predicted timestep to guarantee a range of $[0,1]$.
To condition $\mathcal{P}_\phi$ on the target bitrate, we expand $\lambda$ to the same spatial dimension as the latent sample and concatenate them along the channel dimension before processing with the parameter estimation network.

\subsection{Entropy Coding}
We use a joint contextual and hierarchical entropy model to encode the quantized latent to a bitstream.
Such models have been extensively researched in the literature~\cite{balle2018Variational, minnen2018Joint, qian2022Entroformer}.
We select Entroformer~\cite{qian2022Entroformer} as our entropy model as it yields the best performance in our experiments.
It consists of a transformer-based hyperprior and bidirectional context model to estimate the latent's distribution $P(\mathbf{\hat{z}})$, which is used to encode it to a bitstream via arithmetic encoding~\cite{pasco1976source}.

\begin{algorithm}[t]
    \scriptsize
    \caption{Encoding and Decoding process}
    \begin{algorithmic}
        \State Given: image $\mathbf{x}$
        \State $\mathbf{y} \gets \mathcal{E}_{vae}(\mathbf{x})$
        \State $\gamma, t \gets \mathcal{P}_{\phi}(\mathbf{y})$
        \State $\mathbf{\hat{z}} \gets \mathcal{Q} (\mathbf{y}, \gamma)$
        \State bitstream $\longleftrightarrow \mathbf{\hat{z}}$
        \Comment{Entropy code using $P(\mathbf{\hat{z}})$}
        \State $\mathbf{\hat{y}}_t \gets \mathcal{Q}^{-1}(\mathbf{\hat{z}},\gamma)$
        \For{$n = t$~\textbf{to}~$1$}
        \State $\mathbf{\hat{y}}_{n-1} \gets \mathcal{M}_{\theta} (\mathbf{\hat{y}}_n, n)$
        \Comment{Eq.~\ref{eq:diffusion}}
        \EndFor
        \State $\mathbf{\hat{x}} \gets \mathcal{D}_{vae}(\mathbf{\hat{y}}_0)$
    \end{algorithmic}
    \label{alg:inference}
\end{algorithm}

\subsection{Optimization}
\label{subsec:optimization}
Following Eq.~\ref{eq:rd-loss}, we jointly optimize the tradeoff between the estimated coding length of the bitstream and the quality of the reconstruction:
\begin{equation}
    \mathcal{L}~=-\log_2 P(\mathbf{\hat{z}}) + \lambda \parallel \mathbf{x} - \mathbf{\hat{x}}\parallel ^2 _2.
\end{equation}

We train our model on the \textbf{Vimeo-90k}~\cite{xue2019video} dataset and randomly crop the images to 256$\times$256px in each epoch. Our model is optimized for 300,000 steps with learning rate $1\text{e-}4$. We randomly sample $\lambda \in [1,5,10,20]$ at each gradient update to train for multiple target bitrates within a single model.

While our main motivation was to utilize foundation models without significant modification, we do make minor adjustments in our pipeline to allow for optimization of trainable modules upstream.
During training, it is prohibitively expensive to backpropagate the gradient through multiple passes of the diffusion model as it runs during DDIM~\cite{song2021Denoising} sampling.
Therefore, we perform only one DDIM sampling iteration and directly use $\mathbf{\tilde{x}}_0$ as the fully denoised data~(see~\cref{eq:denoise} and~Appendix B).
For the low timestep range our model operates in, we observe that the difference between $\mathbf{\tilde{x}}_0$ and the true fully denoised data $\mathbf{x}_0$ is minimal and sufficient for the optimization of the parameter estimation module.
At inference time, we perform the standard iterative DDIM process.
Additionally, during the diffusion sampling process, the timestep $t$ is used to index an array of required precomputed values~(\eg,~variance schedule $\alpha_t$).
This discretization prevents optimization of the parameter estimation network. 
Therefore, we implement continuous functions for each required value and evaluate them with the predicted timestep during training.

\section{Experiments}
\label{sec:experiments}

We compare our method to state-of-the-art generative and diffusion-based image compression codecs via objective metrics and a subjective user study.
Fig.~\ref{fig:visual_comparison} also shows qualitative comparisons of ours to different methods.

\subsubsection{Datasets.}
We conduct experiments on the following datasets: i)~\textbf{Kodak}~\cite{kodak1993PhotoCD}, which consists of 24 images 768$\times$512px~(or inverse).
ii)~the~\textbf{CLIC2022}~\cite{2022Challenge} test set, which contains 30 high-resolution images resized such that the longer side is 2048px.
We center crop each image to 768$\times$768px due to the high memory consumption required to process large images with Stable Diffusion;
and iii)~\textbf{MS-COCO 30k}, which has recently been used for evaluating the realism of the reconstruction of compression methods~\cite{agustsson2023MultiRealism, hoogeboom2023HighFidelity}.
The dataset is preprocessed as stated in~\cite{agustsson2023MultiRealism}, resulting in 30,000 images of size 256$\times$256px each.

\subsubsection{Metrics.}

\begin{figure*}[t]
    \centering
    \begin{subfigure}{\linewidth}
        \includegraphics[trim=0 0 0 30,clip,width=\linewidth]{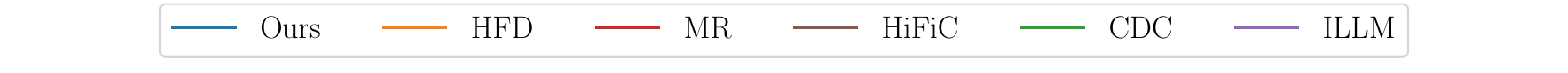}
    \end{subfigure}
    \begin{subfigure}{0.29\linewidth}
        \centering
        \includegraphics[trim=0 -40 0 -40,clip,width=\linewidth]{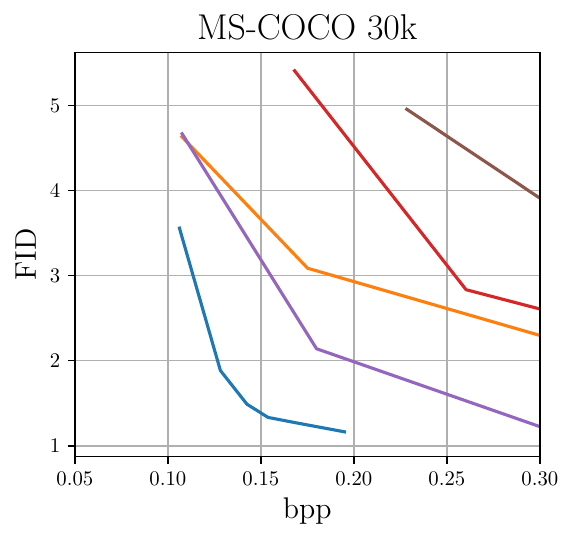}
        \caption{rate-realism}
        \label{subfig:rd-realism}
    \end{subfigure}
    \begin{subfigure}{0.7\linewidth}
        \centering
        \includegraphics[width=\linewidth]{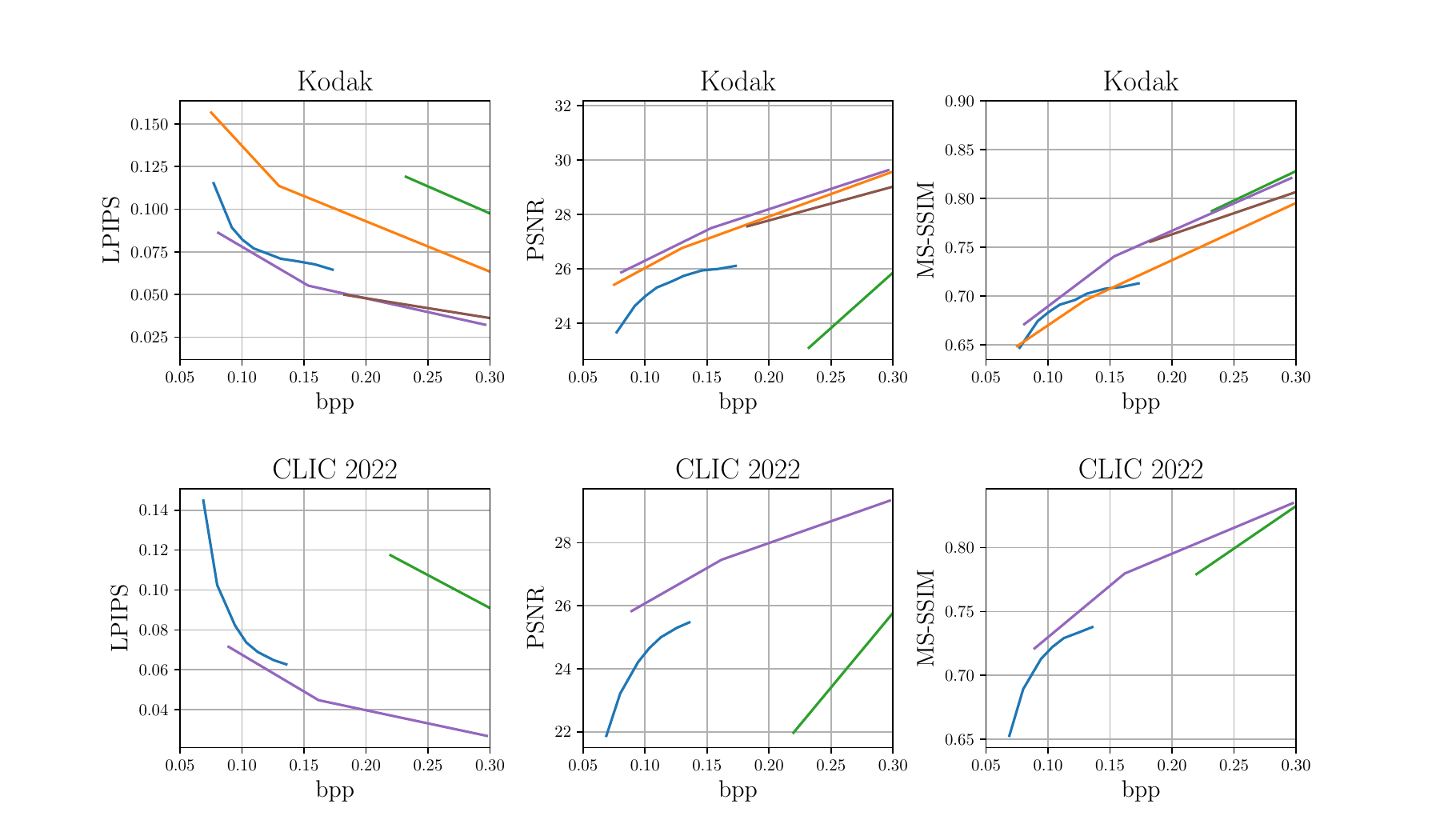}
        \caption{rate-distortion}
        \label{subfig:rd-distortion}
    \end{subfigure}
    \caption{Quantitative comparison of our method with other baselines. We outperform all methods in (a)~rate-realism while remaining competitive with the best performing generative codecs in (b)~pixel-wise distortion metrics.}
    \label{fig:quant_results}
\end{figure*}

We evaluate our proposal and baselines using: \textbf{PSNR} and \textbf{MS-SSIM}, as measures of pixel-wise distortion; \textbf{LPIPS}, as a more perceptually-oriented distortion metric; and \textbf{FID}~\cite{heusel2017gans}, to evaluate the realism of the reconstructed images.
FID measures the similarity between the distributions of source and distorted images, and thus has been widely used as a measure of realism~\cite{mentzer2020HighFidelity,agustsson2023MultiRealism}, particularly for image generation tasks.
Since FID requires a higher number of samples of both source and distorted images, we focus the FID comparison only on MS-COCO 30k.

\subsubsection{Baselines.} We compare our method against GAN- and diffusion-based image codecs.
\textbf{HiFiC}~\cite{mentzer2020HighFidelity} is a well-known generative neural image compression codec and remains a strong GAN-based baseline, while the recently proposed \textbf{ILLM}~\cite{muckley2023Improving} improves upon the HiFiC architecture and is available at more comparable bitrates to ours.
For diffusion-based approaches, we use \textbf{CDC}~\cite{yang2023Lossy} and \textbf{HFD}~\cite{hoogeboom2023HighFidelity}, the only two other practical works in this field.
Both HFD and Careil~\etal~\cite{careil2023image} are trained on large proprietary datasets and therefore cannot be reproduced for comparison. For this reason, we compare to HFD only on the Kodak dataset and FID score on MS-COCO 30k as these are the only results available, and cannot compare to Careil~\etal.
For all other methods, we use released pretrained model weights and run with default parameters.
However, for CDC we increase the number of denoising sampling steps to 1000 to produce higher quality reconstructions.

\subsubsection{User study.} 
\label{subsec:user-study}
As qualitative metrics often fail to capture the perceptual quality of image reconstructions~\cite{blau2019Rethinking, stein2023Exposing}, we further perform a user study to assess the visual quality of our results.
The study is set up as a two-alternative forced choice~(2AFC), where each participant is shown the source image and reconstructions from two methods and is asked to choose the reconstruction they prefer.
We select 10 samples from the Kodak dataset with the smallest difference in bitrate between our method and the other generative model baselines, namely CDC, ILLM, and HFD.
\footnote{In a pilot study, we have also considered to include BPG and HiFiC.
However, since it was clear that they always performed worse than other methods in our target bitrate range, and to avoid fatigue with long rating sessions~(we target a session time around 15--20min) we removed them from the study.}
Thus, in a session, a participant is requested to do 60 pairwise comparisons.
Each sample is center-cropped to 512$\times$512px so that all images being compared are shown side-by-side at native resolution~(\ie, without resampling).
\footnote{The images used in the study and respective bitrates can be found in the Supplementary Materials.}
Participants can freely zoom and pan the images in a synchronized way.
However, because the methods locally provide different types of reconstructions, we ask the participants to inspect the images in their entirety before rating.

Following~\cite{mentzer2020HighFidelity}, we use the Elo~\cite{glickman1995Comprehensive} rating system to rank the methods. Elo matches can be organized into tournaments, where ranking updates are applied only at the end of the tournament. We perform two separate experiments, where a tournament is considered to be \emph{1.}~a single comparison or \emph{2.}~all image comparisons from the same user.
As Elo scores depend on game order, a Monte Carlo simulation is performed over 10,000 iterations, and we report the median score for each method.

\subsection{Results}
\label{subsec:results}

\subsubsection{Quantitative results.} Fig.~\ref{fig:quant_results} shows the rate-distortion~(as measured by PSNR, MS-SSIM, and LPIPS) and rate-realism~(as measured by FID) curves of our methods and baselines.
Our method sets a new state-of-the-art in realism of reconstructed images, outperforming all baselines in FID-bitrate curves.
In some distortion metrics~(namely, LPIPS and MS-SSIM), we outperform all diffusion-based codecs while remaining competitive with the highest-performing generative codecs.
As expected, our method and other generative methods suffer when measured in PSNR as we favor perceptually pleasing reconstructions instead of exact replication of detail~(see Sec.~\ref{sec:nic}). 

\subsubsection{User study.}
\begin{figure}[t]
    \centering
    \begin{subfigure}{0.45\linewidth}
        \includegraphics[width=\linewidth]{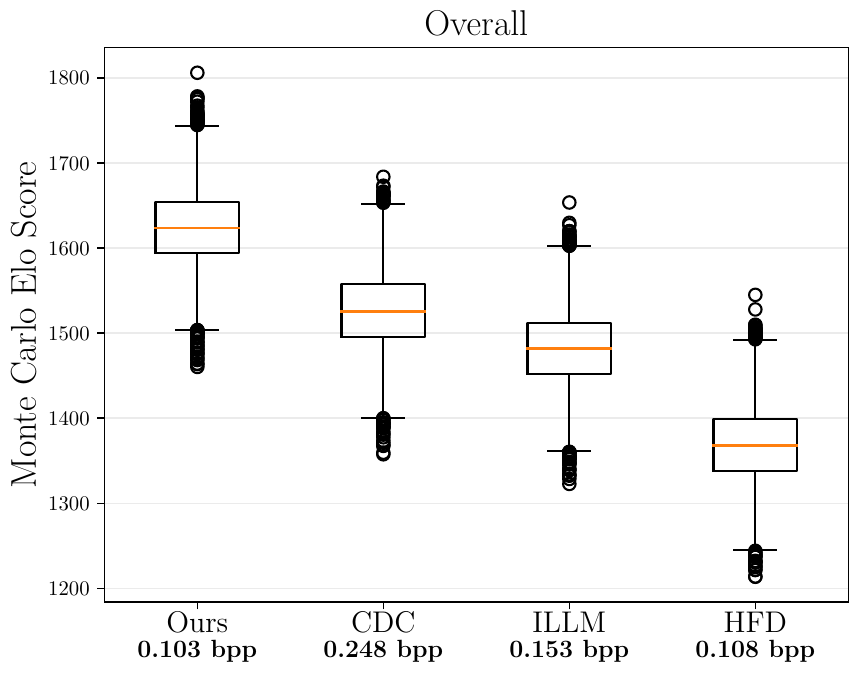}
    \end{subfigure}
    \begin{subfigure}{0.45\linewidth}
        \includegraphics[width=\linewidth]{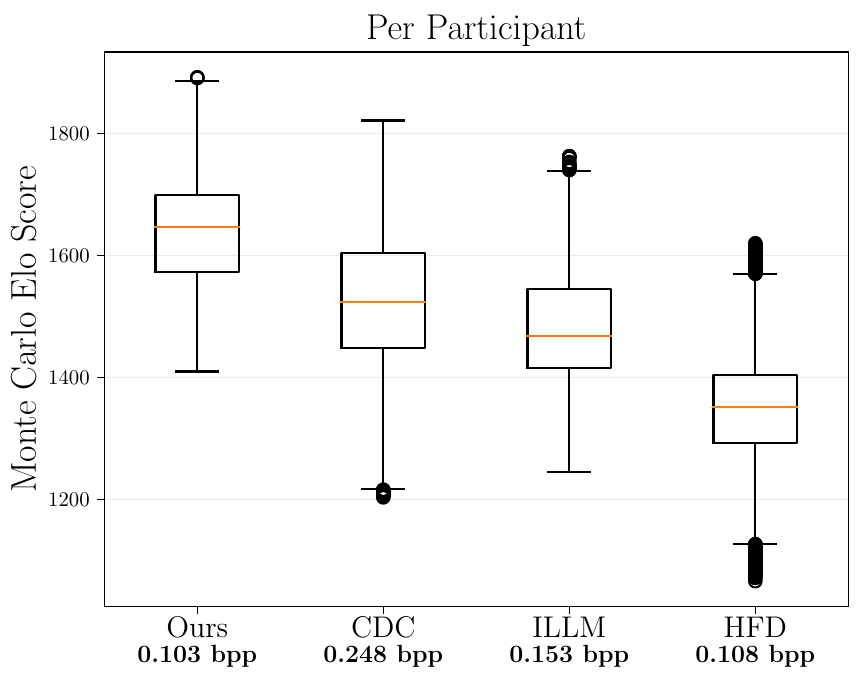}
    \end{subfigure}
    \caption{Computed Elo ratings from the user study with Elo tournaments set for each comparison (left) or for each participant (right). Higher is better. The box extends to the first and third quartiles and the whiskers $1.5\times IQR$ further.}
    \label{fig:user_study}
\end{figure}

Fig.~\ref{fig:user_study} shows the outcome of our user study.
The methods are ordered by human preference according to Elo scores.
The average bitrate of the images for each model are shown below the name of the methods.
As can be seen in the Elo scores, our method significantly outperforms all the others, even compared to CDC, which uses on average double the bits of our method. This remains true regardless of Elo tournament strategy used.

\subsubsection{Visual results.} Fig.~\ref{fig:visual_comparison} qualitatively compares our method to generative neural image compression methods. 
Our approach can consistently reconstruct fine details and plausible textures while maintaining high realism.
HFD often synthesizes incorrect content~(door in row 1, flowers in row 5, and mural in row 6) or produces smooth reconstructions~(flower petal in row 2, face in row 3, red barn in row 4, and face and hat in row 7).
CDC and ILLM introduce unnatural blurry or high-frequency generative artifacts~(flower bud in row 2 and tree in row 4) even in cases where they use 2x the bitrate of our method.

\def\picwidth{0.92\linewidth}
\begin{figure*}[ht!]
\begin{subfigure}{\textwidth}
    \centering
    \includegraphics[width=\picwidth]{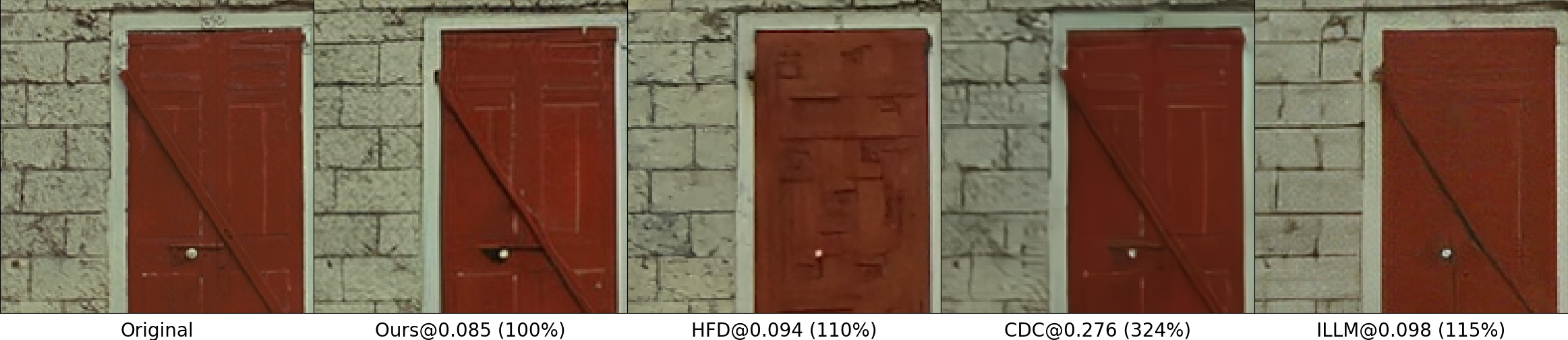}
\end{subfigure}
\begin{subfigure}{\textwidth}
    \centering
    \includegraphics[width=\picwidth]{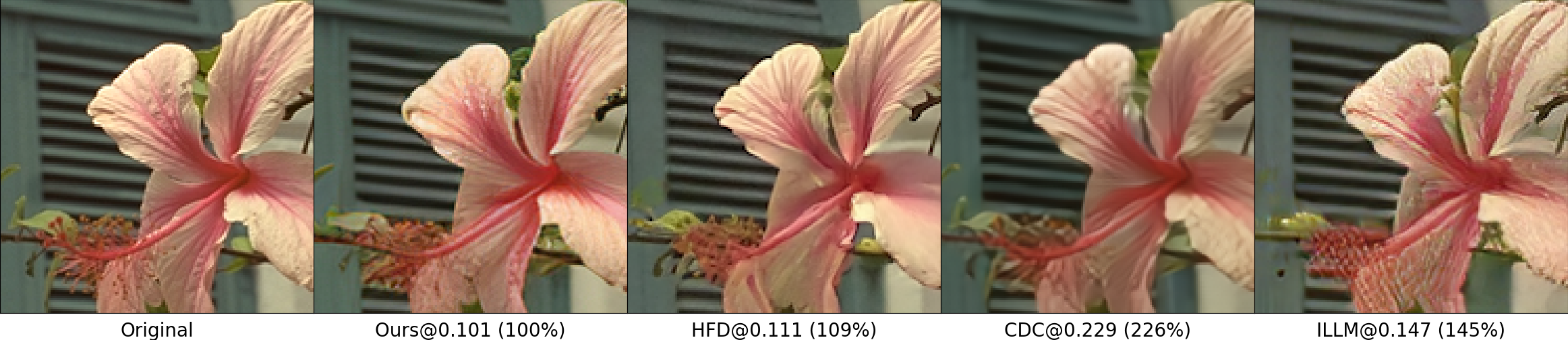}
\end{subfigure}
\begin{subfigure}{\textwidth}
    \centering
    \includegraphics[width=\picwidth]{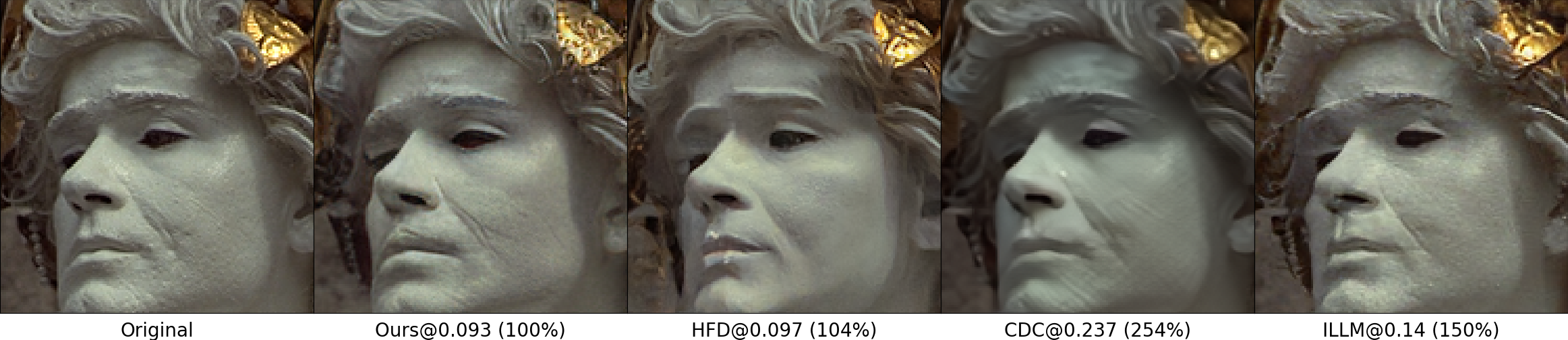}
\end{subfigure}
\begin{subfigure}{\textwidth}
    \centering
    \includegraphics[width=\picwidth]{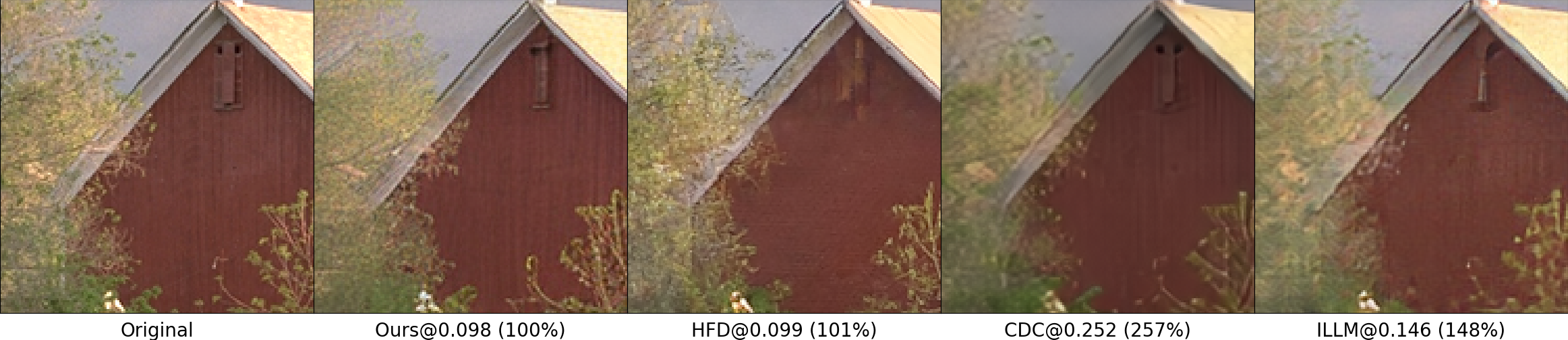}
\end{subfigure}
\begin{subfigure}{\textwidth}
    \centering
    \includegraphics[width=\picwidth]{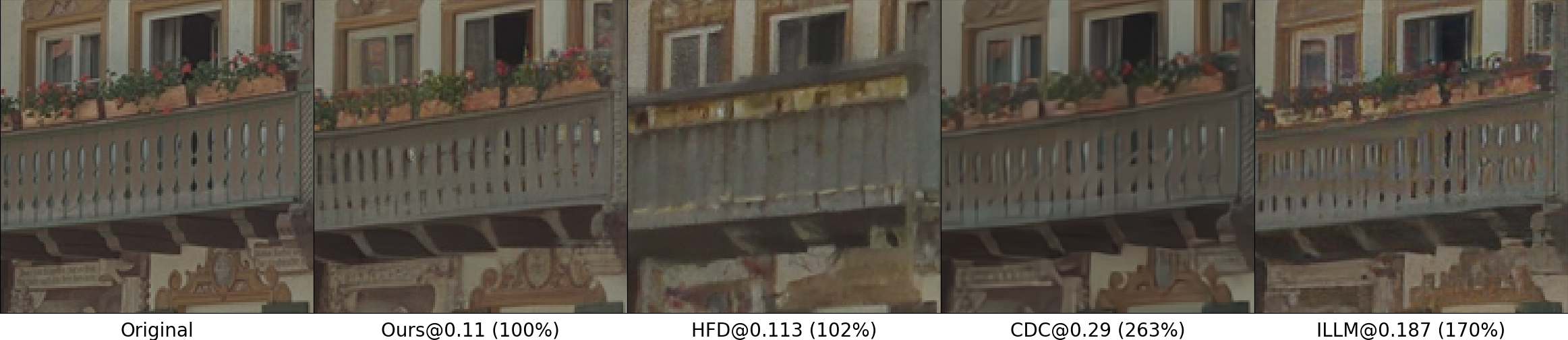}
\end{subfigure}
\begin{subfigure}{\textwidth}
    \centering
    \includegraphics[width=\picwidth]{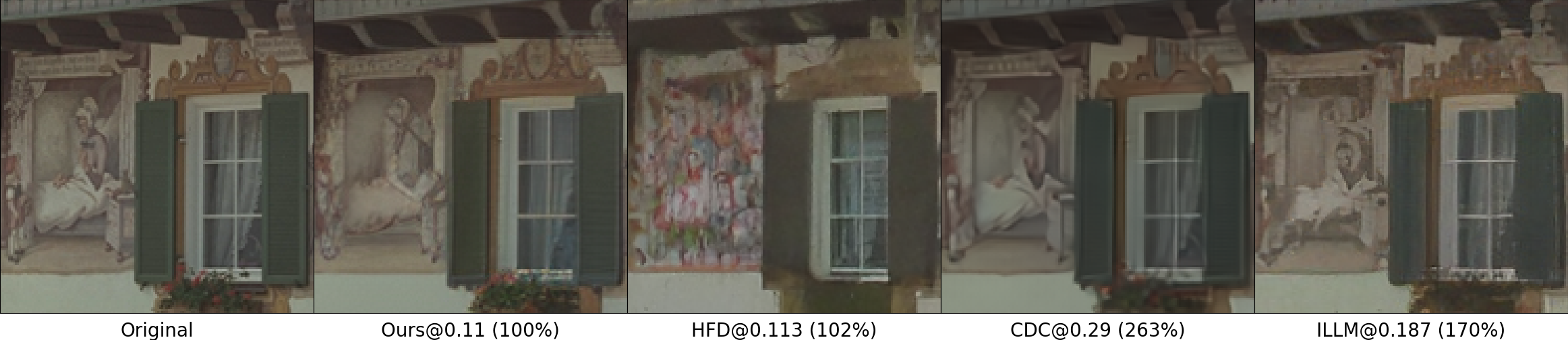}
\end{subfigure}
\caption{Qualitative comparison of our method to the baselines. Images are labeled as Method@bpp (bpp is also shown as a percentage of our method). Best viewed digitally.}
\label{fig:visual_comparison}
\end{figure*}

\subsubsection{Complexity.}
To assess the practicality and efficiency of our method, we compare the runtime and size of our model with CDC and ILLM.
We report the average encoding/decoding time (excluding entropy coding) for all images from the Kodak dataset.
All benchmarks were performed on an NVIDIA RTX 3090 GPU.
Our method processes an image in 3.49 seconds, nearly twice as fast as CDC, which requires 6.87 seconds.
ILLM processes an image in 0.27 seconds.
However, it is important to note that diffusion-based methods are in general slower than other codecs due to their iterative denoising nature.
Due to the Stable Diffusion backbone, our method is more complex than CDC~(1.3B vs. 53.6M parameters, respectively), while ILLM contains 181.5M parameters.
However, the large majority of our parameters come from the diffusion backbone.
Our trained modules~(\eg, $\mathcal{P}_{\phi}$ and the entropy model) contain only 36M parameters.

Reducing the computational burden of diffusion models is an active research area~\cite{hoogeboom2023Simple, jabri2023Scalable}, parallel to ours.
Our method is fundamentally independent of the chosen foundation model, thus advances on reducing the complexity of Stable Diffusion can ultimately also improve our proposal.

\subsubsection{Limitations.}
Similar to other generative approaches, our method can discard certain image features while synthesizing similar information at the receiver side.
In specific cases, however, this might result in inaccurate reconstruction, such as bending straight lines or warping the boundary of small objects.
These are well-known issues of the foundation model we build upon, which can be attributed to the relatively low feature dimension of its VAE.
Despite this, our generated content is still closer to the original content than the other diffusion compression methods, as confirmed by our subjective study, and can be qualitatively compared on Fig.~\ref{fig:visual_comparison}.

\paragraph{Ethical concerns.}
A core challenge of generative machine learning is the misgeneration of content.
Specifically at very low bitrates, identities, text, or lower-level content can vary from the original image, and thus may raise ethical concerns in specific scenarios.

\section{Conclusion}
\label{sec:conclusion}

Via our proposed novel lossy image compression codec based on foundation latent diffusion, we produce realistic image reconstructions at low to very low bitrates, outperforming previous generative codecs in both perceptual metrics and subjective user preference.
By combining the denoising capability of diffusion models with the inherent characteristics of quantization noise, our method predicts the ideal number of denoising steps to produce perceptually pleasing reconstructions over a range of bitrates with a single model.
Our formulation has faster decoding time than previous diffusion codecs and, due to reusing a foundation model backbone, a much lower training budget.
Potential future work includes the integration of more efficient backbone models~\cite{hoogeboom2023Simple, jabri2023Scalable} and the support for user control to navigate the rate-distortion-realism trade-off.

\vfill

\bibliographystyle{splncs04}
\bibliography{main}

\end{document}